\documentclass[10pt, conference]{IEEEtran}
\IEEEoverridecommandlockouts
\usepackage{cite}
\usepackage{amsmath,amssymb,amsfonts}
\usepackage{algorithmic}
\usepackage{graphicx}
\usepackage{textcomp}
\usepackage{xcolor}
\usepackage{listings}
\usepackage{url}
\newcommand{\ctext}[1]{\raise0.2ex\hbox{\textcircled{\scriptsize{#1}}}}
\def\BibTeX{{\rm B\kern-.05em{\sc i\kern-.025em b}\kern-.08em
    T\kern-.1667em\lower.7ex\hbox{E}\kern-.125emX}}

\makeatletter 
\newcommand{\linebreakand}{%
  \end{@IEEEauthorhalign}
  \hfill\mbox{}\par
  \mbox{}\hfill\begin{@IEEEauthorhalign}
}
\makeatother 

\begin{document}

\title{
  Monitoring Cascading Changes of Resources\\in the Kubernetes Control Plane
}

\author{
\IEEEauthorblockN{Tomoyuki EHIRA}
\IEEEauthorblockA{
\textit{Kyoto University}\\
Kyoto, Japan\\
ehira@net.ist.i.kyoto-u.ac.jp}
\and
\IEEEauthorblockN{Daisuke KOTANI}
\IEEEauthorblockA{
\textit{Kyoto University}\\
Kyoto, Japan\\
kotani@media.kyoto-u.ac.jp}
\and
\IEEEauthorblockN{Hiroki SHIROKURA}
\IEEEauthorblockA{
\textit{LINE Corporation}\\
Tokyo, Japan\\
hiroki.shirokura@linecorp.com}
\and
\linebreakand 
\IEEEauthorblockN{Hirofumi ICHIHARA}
\IEEEauthorblockA{
\textit{LINE Corporation}\\
Tokyo, Japan\\
hirofumi.ichihara@linecorp.com}
\and
\IEEEauthorblockN{Yasuo OKABE}
\IEEEauthorblockA{
\textit{Kyoto University}\\
Kyoto, Japan\\
okabe@i.kyoto-u.ac.jp}
}

\maketitle

\begin{abstract}

Kubernetes is a container management system that has many automated functionalities.
Those functionalities are managed by configuring objects and resources in the control plane.
Since most objects change their state depending on other objects' states, 
a change propagates to other objects in a chain.
As cluster availability is influenced by the time required for these cascading changes, it is essential to make the propagations measurable and shed light on the behavior of the Kubernetes control plane.
However, it is not easy because each object constantly monitors other objects' status and acts autonomously in response to their changes to play its role.
In this paper, we propose a measurement system that outputs objects' change logs published from the API server in the control plane and assists in analyzing the time of cascading changes between objects by utilizing the relationships among resources.
With a practical change scenario, our system is confirmed that it can measure change propagation times
within a cascading change.
Also, measurements on the system itself showed it has a small CPU and memory footprint.

\end{abstract}


\section{Introduction}

Lightweight and high-speed container technology has become popular, and novel methods such as microservices\cite{Dragoni2017MicroservicesYT}, in which a system is built by orchestrating many containers, are now used to provide services.
Kubernetes \cite{kubernetes,burns2016borg} is the most popular one among container orchestration systems.

On Kubernetes clusters, cluster administrators set up resources and objects by specifying their desired states declaratively.
Each object represents the desired state and the current state of cluster functions, and a resource is a group of the same kind of objects.
Controllers are components that control the objects of each resource and continuously change the current state toward the desired state (reconciliation loop).
Such mechanisms enable autonomous convergence to the desired state when the desired state changes (e.g., an administrator changes cluster settings) or when the current status changes (e.g., a machine fails accidentally or resource usage exceeds a threshold).
Controllers update their objects' status depending on other objects' status, so one change to an object would cause cascading changes and propagate throughout the cluster.
In this way, the control plane is responsible for controlling a Kubernetes cluster.

One of the important metrics on Kubernetes cluster monitoring is the time from a change of objects because of external factors including updates of the desired state by administrators and the actual state by nodes in the cluster, to the completion of cascading changes, that is, the actual status of related objects are converged to the desired status.
The time required for the change propagation directly impacts the quality of service, for example, the tolerance of the system to sudden increases in user requests is affected by the time from the change of the object representing the desired number of containers to the change of the object representing the actual number of containers running correctly.
In order to detect bottlenecks and improve performance, a system that visualizes and quantifies the propagation time is required.

However, this is not easy primarily due to two factors.
The first reason is that the change process of individual objects is carried out autonomously in the control plane so that the progress of the process is not explicitly expressed.
The sequence of object changes is not specified procedurally, but rather, after setting up an object, various objects are changed autonomously to adapt to the change. So unless the administrator explicitly obtains the objects' status, it is impossible to know what state each object is in and when all relevant changes have been completed.
Second, it is impossible to know without prior knowledge which other objects will be affected by a change in one object.
Each controller independently gets the current objects' state and processes to make them the desired states, so when focusing on each object, it is impossible to know which objects are affected by a change without knowing the logic of controllers.

In this paper, we propose a system that monitors cascading changes between dependent resources utilizing the API server \cite{apiserver}, which accepts object changes and measures the time required for cascading changes between resources or objects using our system.
This system consists of a logging component that outputs object change messages sent from the API server into a log file as an entry with a timestamp when it is received and a log-analyzing component that calculates the time taken for the change to propagate and aggregates it by linking the entries based on the dependency information and taking the difference between the timestamps in the logs.
The advantage of this method is that it does not modify existing components of the control plane so that the measurements can be performed on a Kubernetes cluster in production rather than on a dedicated experimental Kubernetes cluster.

The contributions of this research are as follows:
\begin{itemize}
  \item We highlight the importance of the metric ``time required for configuration changes to propagate,'' which has not received much attention so far, and propose a framework for measuring change propagation times in Kubernetes.
  \item We explain why it is difficult to measure the time required for changes in Kubernetes, where the state of objects is controlled autonomously in the control plane.
  \item We define dependencies as relationships between resources or objects in some specific cases.
  \item We implement a low-overhead system to measure various change propagation times and confirm with general scenarios that the time required for state changes can be visualized and quantified.
\end{itemize}

\section{Background}\label{sec:background}
\subsection{The Kubernetes Control Plane}

Kubernetes is logically divided into two parts; the data plane, where containers run, and the control plane, which manages containers running in the data plane and the cluster as a whole in order to perform various functions.

The control plane maintains the state of the cluster as many sets of objects, and a resource is a set of objects with the same functionality.
Typical resources provided by the default are as follows.
Pod is a logical machine in which an application container runs.
ReplicaSet represents a set of Pods with the same configuration and ensures that a specified number of Pods are running at any given time.
If the current number of Pods does not match the specified number of replicas for some reason, Pods are created or deleted.
Service exposes a set of pods as a network service.
Endpoints hold a list of Pod endpoints selected by the Service object.
In addition to these standard resources, additional custom resources can be defined to extend the cluster's automaticity.

\begin{figure}[!t]
  \centering
  \includegraphics[keepaspectratio, width=0.95\hsize]
  {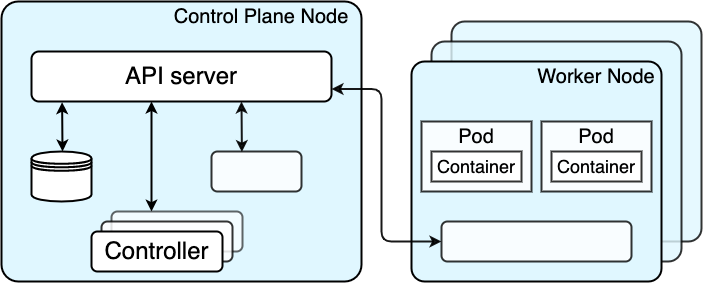}
  \caption{Kubernetes architecture}
  \label{fig:control-plane}
\end{figure}

Fig. \ref{fig:control-plane} shows a schematic representation of a Kubernetes cluster.
The API server listens to object changes from administrators and control plane components. When an object is modified, it notifies the entire controller that has subscribed to the object's modification for that resource of the change.
Controllers are responsible for changing the objects of the corresponding resource so that the objects are in the desired state. The controllers of each resource obtain the current state of the corresponding resource and the objects of dependent resources through the API server's subscription function. If the current state differs from the desired state, they change it to the desired one. This process in a controller is called a reconciliation loop.

\subsection{Related Work}

Distributed tracing is a technique for tracking how user requests propagate between services (applications running in containers) in the data plane.
The annotation-based method \cite{Dapper-36356, Fonseca2007XTraceAP, 10.5555/647883.738238, 1021256}
allows detailed tracking of each request from the user.
This is accomplished by assigning a unique identifier to each user request, recording it with a timestamp when each service receives it, and assigning the same identifier when invoking other services.
Although there are differences between the control plane and the data plane, distributed tracing is similar to the purpose of this paper, which is to monitor and measure propagation.
However, when tracking the behavior (changes) of control planes, many factors cause changes, like external factors such as machine failures and internal factors such as configuration changes by administrators, making it challenging to identify the source of the change and give it an identifier.
In addition, since each control plane component operates autonomously and does not invoke other components, it is hard to measure change propagation by propagating annotations along with the change.

With a similar objective, K-Bench\cite{k-bench} and ClusterLoader\cite{clusterloader} measure the time required for transitions between some specific states of a pod object.
However, these can only measure in a limited number of scenarios, and they do not support monitoring and measurement of change propagations between other existing resources or custom resources defined by the cluster administrator.

kube-state-metrics\cite{kube-state-metrics} uses the API server to output the detailed state of objects on the cluster and visualize the current cluster state. However, it requires manual coding to output information for each custom resource (e.g., third-party plugins).

Like Kubernetes, the Internet has an auto-recovery mechanism that automatically updates routes to handle partial failures due to faults.
As with cascading changes, it is difficult to determine where the change started and when the route propagated.
To address this problem, the time required to propagate a change (route) and the time required for the route to converge is measured by measuring the time it takes for the route to change to the IP prefix at a remote location (BGP beacon) for an IP prefix that is periodically advertised at a router at a specific location. Studies by Mao et al. \cite{10.1145/948205.948207} and Feldmann et al. \cite{10.1145/1015467.1015491} attempt to determine the cause and location of routing changes.
This approach differs in that it only obtains information primarily from the sending router and the receiving router under its control and can only measure the time between the ends, whereas, in a Kubernetes cluster, we want to measure the time of propagation between each object along the way. The problem setup is also different in that in Kubernetes, the state of all objects can be obtained through the API server.

\section{Proposed System}\label{sec:system}
We describe the proposed measurement system and the analysis method of cascading changes using this system.

\subsection{System Overview}\label{subsec:system-overview}
This system subscribes to changes in target resources and outputs the change information to a log, taking advantage of the fact that all changes in the cluster are processed in the API server. Then, administrators analyze, aggregate, and visualize the changes in the logs using the aggregator, a library that analyzes the logs using dependency data given from the cluster admin. There is also a component to measure the changes by modifying objects according to a given scenario.
The design principles of this system are as follows.
\begin{itemize}
  \item Do not modify the existing Kubernetes components so that the system can be deployed in production clusters. 
  \item Keep a log with times of changes and the state of the object at that point for later flexible analysis.
  \item Do not require cumbersome codings in logging.
  \item Introduce mechanisms to facilitate analysis of change propagation between dependent objects (resources) in order to measure valuable time from the logs, rather than just logging disorderly about the state of objects.
  \item Ensure that the intervention of the monitoring system imposes no significant load on the Kubernetes system. Also, avoid causing additional delays in the propagation time.
\end{itemize}

\subsection{System Architecture}\label{sebsec:system-architecture}
\begin{figure}[!t]
  \centering
  \includegraphics[keepaspectratio, width=0.90\hsize]
  {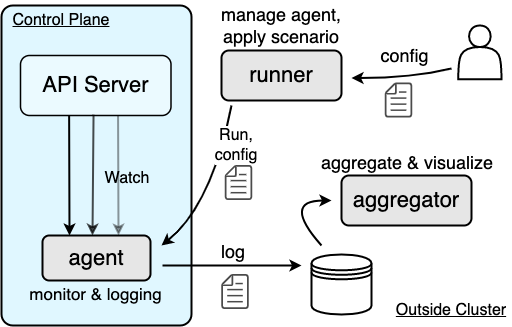}
  \caption{Architecture of the proposed system}
  \label{fig:system-architecture}
\end{figure}

Fig. \ref{fig:system-architecture} shows the architecture of the proposed system.

The proposed system consists of three components; agent, aggregator, and runner. The functions and implementations of each component are as follows.

\begin{table}[!t]
  \caption{Fields of agent outputs}
  \label{tab:log_format}
  \centering
  \begin{tabular}{ll}
    \hline
    Field Name   & Description                              \\
    \hline
    \textit{Time}         & Time that notification had arrived from API server \\
    \textit{Op}           & Kind of change (\textit{Add}, \textit{Update} or \textit{Delete}) \\
    \textit{Obj}          & Dump of object  \\
    \textit{OldObj}       & Dump of old object before update        \\
    \hline
  \end{tabular}
\end{table}

\begin{figure}[!t]
  \centering
  \caption{agent config (partially omitted)}
  \includegraphics[keepaspectratio, width=0.99\hsize]
  {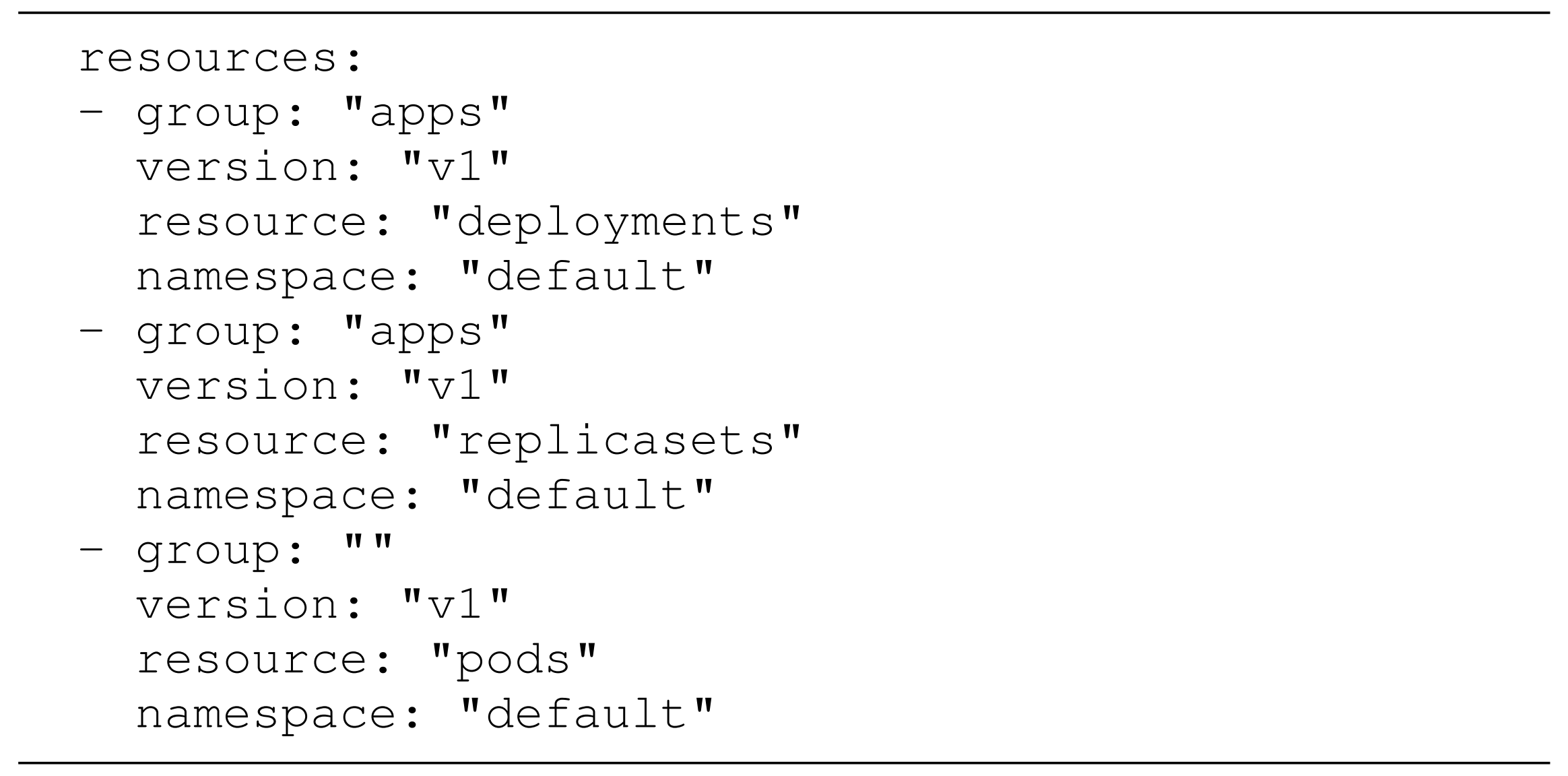}
  \label{code:agent-config}
\end{figure}

\noindent
\textbf{agent}\quad
runs on a cluster, monitors and logs changes to objects in the specified resource.
Fig. \ref{code:agent-config} is an example of agent configuration and the fields of the log output by the agent are shown in Table \ref{tab:log_format}. When it receives the notification about a modified object from the API server, it immediately records the current time as a timestamp (\textit{Time}),
the operation to the object (add, update, or delete) (\textit{Op}),
the current status of the object (\textit{Obj}), and in the case of an update, the previous status of the object (\textit{OldObj}).

\noindent
\textbf{aggregator}\quad
is a library and a visualization tool used to analyze logs output by agent.
It extracts the time of change for each resource from the log and quantifies and visualizes the time required for change propagation based on given dependencies between objects.

\noindent
\textbf{runner}\quad
is a component that invokes object changes according to a scenario by using agent.
The measurement in Section \ref{sec:measure-changes} uses runner. 
First, copy and save all the parameters of the measurement.
After that, we specify the resource to be observed, start the agent, and make changes according to the scenario.
When all changes have been finished, the log file is saved and the process ends.

\subsection{How to Analyze}
The agent outputs a large number of logs, and it is necessary to extract and analyze the necessary information from these logs.
This section describes how to measure the time required for a chain of changes from the logs, including an explanation of dependencies in Kubernetes.

The term ``dependency'' has two meanings; dependency between resources and dependency between objects.
When a pod depends on a ReplicaSet, it is a relationship between resources, which is used conceptually, whereas when a change in the settings of a ReplicaSet causes a Pod to be replicated, it is a dependency between objects where changes are actually propagated.

The dependencies between objects may have any correspondences depending on controller implementations.
From the correspondences used in the standard Kubernetes resources, we found three dependencies: owner relationships, name-prefix relationships, and label relationships.
The owner relationships are relationships that explicitly refer to the dependant object in the OwnerReferences field in an object's metadata, such as between Deployment and ReplicaSet, and between ReplicaSet and Pod.
The name-prefix relationships are relationships in which the name of the dependent object is taken over as a prefix of their names.
The label relationships are relationships that depend on the state of an object with a specified label, such as the relationship between Pod and Endpoints, where Endpoints holds the endpoints of a Pod with a specified label.
Labels are used to specify a set of objects to be managed together.

By providing information on dependencies between resources in the aggregator, it can be converted to dependencies between objects contained in logs, and related logs can be automatically extracted to illustrate the relationship between dependent objects.
The time required for change propagation is measured by subtracting the difference between log timestamps while using such information.

\section{Evaluation}\label{sec:measure-changes}
We measure the time required for a chain of changes between dependent resources using the proposed system.
In Section \ref{subsec:DRP-measure}, we measured the propagation time for creation between Deployment, ReplicaSet, and Pod.
In Section \ref{subsec:overhead}, we measured CPU and memory utilization during the measurement and evaluated the overhead of the proposed system.

\subsection{Environment}\label{subsec:measure-env}
The experimental Kubernetes cluster consists of five VMs and is created using kubeadm \cite{kubeadm}.
One of these machines is a control plane node, where control plane components such as the API server run, and the other four are worker nodes, where application containers run.

\subsection{Measurement among Deployment, ReplicaSet and Pod}
\label{subsec:DRP-measure}

Deployment and ReplicaSet and ReplicaSet and Pod are typical examples of dependent resources in Kubernetes. We measured the time until changes were made to these resources using the implementation. In addition, we observed the variation of propagation time by changing the number of replicas of the Pod.

\subsubsection{Procedure}
The first scenario is ``adding and deleting a Deployment object with $N$ Pods''.

\begin{figure}[!t]
  \centering
  \includegraphics[keepaspectratio, width=0.80\hsize]
  {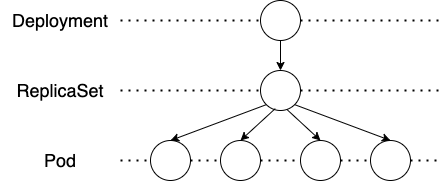}
  \caption{Object dependencies in the first scenario  \ ($N=4$)}
  \label{fig:DRS_1N}
\end{figure}

The schematic diagram of the object dependencies of the first scenario is shown in Fig. \ref{fig:DRS_1N}.
Ten measurements were taken for each case where $N$ is $100$, $200$, and $400$, respectively.

\subsubsection{Analysis and Discussion}
\begin{figure}[!t]
  \centering
  \includegraphics[keepaspectratio, width=0.95\hsize]
  {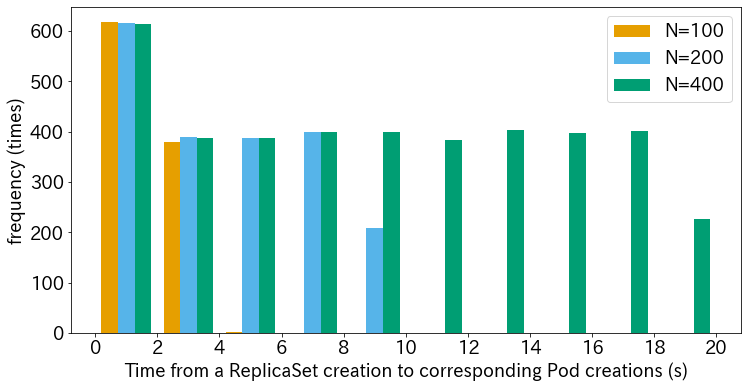}
  \caption{Elapsed times of \textit{Add} from ReplicaSet to Pod (with $N$ Pods)}
  \label{fig:1xN_RS2Pod_Add}
\end{figure}

\begin{figure}[!t]
  \centering
  \includegraphics[keepaspectratio, width=0.95\hsize]
  {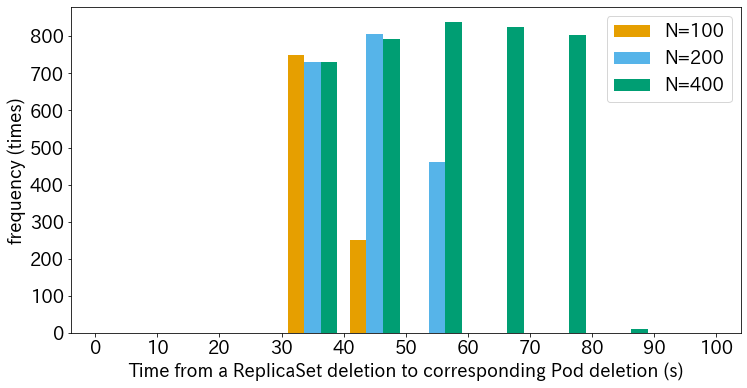}
  \caption{Elapsed times of \textit{Delete} from ReplicaSet to Pod (with $N$ Pods)}
  \label{fig:1xN_RS2Pod_Delete}
\end{figure}

Fig. \ref{fig:1xN_RS2Pod_Add} and \ref{fig:1xN_RS2Pod_Delete} show the combined results of 10 measurements, expressed as a histogram.
Fig. \ref{fig:1xN_RS2Pod_Add} shows the time from when a ReplicaSet object is created until the corresponding Pod object is created,
and Fig. \ref{fig:1xN_RS2Pod_Delete} shows the time from when a ReplicaSet object is deleted until the corresponding Pod object is deleted.
The horizontal axis represents the time slots (ms) and the vertical axis represents the frequency of each time to complete the operation (times).


From Fig. \ref{fig:1xN_RS2Pod_Add}, the number of pods created per time stays constant even if $N$ increases.
This is likely because the ReplicaSet controller creating the Pods limits the speed of Pod creation. 
We can also observe that the time to complete the creation of all pods is roughly proportional to the number of replicas $N$.
And from Fig. \ref{fig:1xN_RS2Pod_Delete}, there is an increase in the number of pods deleted per unit of time as time passes, and the time required to complete the deletion of all pods is roughly proportional to the number of replicas $N$.

\subsection{Measurement of Overhead}\label{subsec:overhead}
The effect of the proposed system on the control plane is evaluated in terms of the CPU and memory utilization of agent. For comparison, we also measured those of the API server at the same time.

\subsubsection{Procedure}
The measurement scenario we use here is the same one in Section \ref{subsec:DRP-measure} ($N = 100$).
The agent was configured to subscribe to three resources; Deployment, ReplicaSet, and Pod. We measured the CPU usage and memory usage of the agent and API server respectively during this period.

\subsubsection{Analysis and Discussion}
\begin{figure}[!t]
  \centering
  \includegraphics[keepaspectratio, width=0.95\hsize]
  {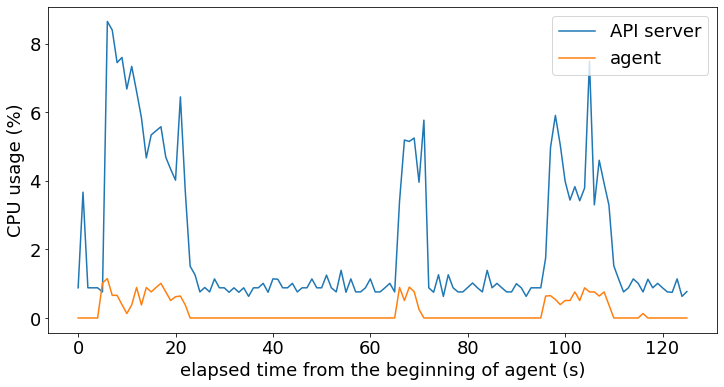}
  \caption{CPU usages of agent and API server}
  \label{fig:cpu-usage-graph}
\end{figure}

\begin{figure}[!t]
  \centering
  \includegraphics[keepaspectratio, width=0.95\hsize]
  {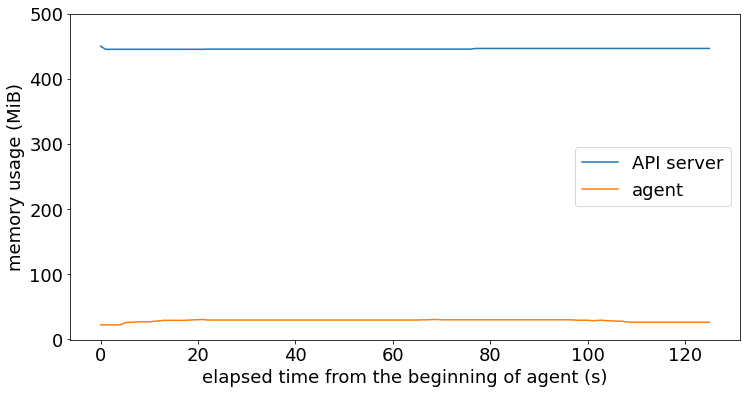}
  \caption{memory usages of agent and API server}
  \label{fig:mem-usage-graph}
\end{figure}

The CPU and memory usage during the measurement is shown in Fig. \ref{fig:cpu-usage-graph} and \ref{fig:mem-usage-graph}.
Both average CPU and memory usage are approximately 10\% or less compared to the API server.
Also, Fig. \ref{fig:cpu-usage-graph} shows that the CPU usage is almost 0\% when there is no change in the monitored resource, and the CPU usage of the agent is smaller than that of the API server at all times during the measurement.
From these results, the impact of this measurement system on the CPU and memory usage of the existing control plane is small.


\section{Discussion}\label{sec:discussion}
Here we discuss our system from several aspects shown in survey papers on cloud system monitoring\cite{Fatema2014ASO,ward2014observing}.

\noindent
\textbf{Scalability}\quad 
Since all object changes are sent from the API server, this system, which uses the API server subscription mechanism, is not affected by an increase in the number of nodes.
We have not quantitatively examined how much this system scales with increases in the number of monitored resources or object change requests, and we will work on this issue in the future.

\noindent
\textbf{Fault tolerance}\quad 
By containerizing the agent itself as an application and running it as a pod, the fault tolerance properties of Kubernetes can be used to automate failure recoveries.
However, changes notified during the restarts cannot be output to the log.

\noindent
\textbf{Time sensitivity}\quad 
In the proposed system, there may be a delay of nanoseconds to milliseconds between the actual state change of an object and the time stamp recorded by the agent, due to internal processing of the controller and change notifications by the API server.
However, this is acceptable for the purpose of identifying major bottlenecks in change propagation.

\noindent
\textbf{Comprehensiveness}\quad 
All objects of resources managed by the cluster, including custom resources that are added later, can be easily change-logged by simply writing the resource name as in Fig. \ref{code:agent-config}.

\section{Conclusion}\label{sec:conclusion}
In this paper, we proposed a system to measure the time required to propagate changes among objects by using the API server mechanism to output logs of object changes and analyze the logs to identify bottlenecks and improve performance. 
The proposed system can be deployed without affecting the operation of clusters. Furthermore, compared to existing systems with fixed types of observable resources and observable state transitions, our system can flexibly measure changes to any resource by just specifying resource names and their relationships in a configuration file.
While handling raw logs allows for flexible analysis, it also makes the analysis work more difficult. However, the proposed system has a mechanism that automatically associates dependent objects with given resource dependencies, making it possible to analyze various types of change propagations efficiently.
We implemented a PoC of the proposed system and quantified how long it takes to create and delete Deployment, ReplicaSet, and Pod resources, and found that the number of Pods created per unit of time by a single ReplicaSet is constant and that the number of deletions per unit of time increases over time.
Furthermore, we evaluated the performance of the implemented system itself and confirmed that both CPU utilization and memory usage were small enough.

In the future, we plan to study not only Kubernetes but also distributed systems with autonomous characteristics and organize requirements for the control plane measurements of such systems and effective monitoring of their control planes.

\bibliographystyle{IEEEtran}
\bibliography{paper,IEEEabrv}
\end{document}